\documentclass[aps,prd,twocolumn,eqsecnum,showpacs,amsmath]{revtex4}

\usepackage{subfigure}
\usepackage[dvips]{color,graphicx}
\usepackage{amsfonts,amssymb,theorem}
\newcommand{\ma}[1]{\mbox{$\mathcal{#1}$}}

{\theorembodyfont{\upshape}
}
{\theorembodyfont{\upshape}
}
{\theorembodyfont{\upshape}
}
{\theorembodyfont{\upshape}
}
{\theorembodyfont{\upshape}
}
{\theorembodyfont{\upshape}
}

\newcommand{\dalm}{\kern1pt\vbox{\hrule height 0.9pt\hbox{\vrule width
0.9pt\hskip 2.5pt\vbox{\vskip 5.5pt}\hskip 3pt\vrule width 0.3pt}\hrule height
0.3pt}\kern1pt}

\begin{document}

\title{
Simple analytic model of wormhole formation
}

\author{Hideki Maeda}
\email{hideki@cecs.cl}


\address{ 
Centro de Estudios Cient\'{\i}ficos (CECS), Arturo Prat 514, Valdivia, Chile
}

\date{\today}

\begin{abstract} 
An explicit and simple solution representing the wormhole formation is presented.
The spacetime is constructed by gluing the Minkowski and Roberts spacetimes at null hypersurfaces in a regular manner.
The parameters in the Roberts solution are required to give the negative kinetic term for the massless scalar field.
Although a curvature singularity appears at the moment of the wormhole formation, it disappears instantaneously.
This instantaneous singularity is weak in the senses of both Tipler and Kr\'{o}lak along radial causal geodesics.
\end{abstract}

\maketitle


\section{Introduction}
Along with black holes, wormholes are intriguing objects in general relativity which have been attracting people even not working in gravitational physics.
A wormhole is locally characterized by a ``throat,'' i.e., a two-dimensional compact spatial surface of minimal area on an achronal hypersurface, connecting some asymptotic regions or infinities.
Wormholes admit the (apparent) superluminal travel as a global effect of the spacetime topology~\cite{visser,superluminal,lobo2007}.
Moreover, they are available to make time machines~\cite{mty1988,timemachine}.
(The readers should refer to~\cite{visser} for a standard textbook and~\cite{lobo2007} for a nice recent review.)

The Morris-Thorne static traversable wormhole connecting two asymptotically flat spacetimes is now a well-known classic in general relativity~\cite{mt1988}.
(Static wormhole metrics were obtained even before Morris and Thorne~\cite{before}.)
It is known that an exotic matter violating the null energy condition is necessary for static traversable wormholes in general relativity~\cite{visser,hv1997,negative}.
This is also a natural consequence of the topological censorship in the asymptotically flat case~\cite{TC}. 
Thus, to construct wormhole solutions with small or even without violation of the energy condition has been a big challenge in wormhole physics~\cite{bd,gb}.
In fact, it was shown that the wormhole spacetime can be constructed with an arbitrarily small amount of matter which violates the averaged null energy condition~\cite{vkd2003}.
This result suggests that the wormhole configuration could be realized in the universe by some quantum effects violating the energy conditions.

Then, a natural question is the stability of the wormhole solutions.
A wormhole could be formed from the gravitational collapse of matter fields possibly violating the energy conditions.
Also, it may be formed by some quantum tunneling effect.
The stability analysis is important in order to clarify the stable stationary configuration of a wormhole.
In the case of the static wormhole solution with thin shells, there exist linearly stable configurations depending on the parameter(s) of the solution~\cite{thin}. 
On the other hand, no stable and analytic wormhole solutions have been reported in the studies of mode analyses and numerical simulations so far~\cite{sh2002,a-p2002,ggs2008,bg2001}.

Independent of the stability of the stationary wormhole configurations, the formation of a wormhole is a highly nontrivial problem because it is a dynamical process of the topology change.
Actually, the dynamical aspects of wormholes have not been well understood so far.
Although there is a lot of static wormhole solutions obtained in the literature, there are few works on the exact model of wormhole formation from the regular initial data.
Because the formation or the growth of a wormhole is essentially a quite complicated dynamical and inhomogeneous process, numerical methods have been often used to study such problems.
In these surroundings, exact analytic models are important to give a transparent picture of the phenomenon.
They become test beds for the future research and should be intensively investigated to complement the numerical works.

The purpose of the present paper is to give a simple analytic model of wormhole formation with a massless ghost scalar field.
The rest of the present paper is constituted as follows.
In the following section, basic equations and a review of the Roberts solution are presented. 
In Sec.~III, we construct our model and study its global structure.
In Sec.~IV, properties of the curvature singularity in the Roberts spacetime are studied.
Concluding remarks and discussions including future prospects are summarized in Sec.~IV.
In Appendix A, the relation between the Roberts and the Gutman-Bespal'ko solutions is explicitly shown.
In Appendix B, the global structure of the Roberts solution for the nonghost case is reviewed.
We adopt the units such that $c=G=1$.
The metric signature convention is taken to be $(-,+,+,+)$, and greek indices run over all spacetime indices.
The conventions of the curvature tensors are 
$[\nabla _\rho ,\nabla_\sigma]V^\mu ={{\cal R}^\mu }_{\nu\rho\sigma}V^\nu$ 
and ${\cal R}_{\mu \nu }:={{\cal R}^\rho }_{\mu \rho \nu }$.


\section{Model and the solution}
We begin with the following action: 
\begin{equation}
S=\int d^4x\sqrt{-g}\left[\frac{1}{16\pi}{\ma R}-\frac12 \epsilon \phi_{,\mu}\phi^{,\mu}\right],
\label{action}
\end{equation}
where $\epsilon=1$ and $-1$ respectively correspond to the real and ghost massless scalar field.
The energy-momentum tensor for a scalar field is given by 
\begin{equation}
T_{\mu\nu}=\epsilon \left(\phi_{,\mu}\phi_{,\nu}-\frac{1}{2}g_{\mu\nu}\phi_{,\rho}\phi^{,\rho}\right).
\label{eq:stress-energy_tensor_of_scalar_field}
\end{equation}
The Einstein equation is 
\begin{equation}
{\ma R}^{\mu}_{~~\nu}=8\pi \epsilon \phi^{,\mu}\phi_{,\nu}, \label{beq}
\end{equation}
while the equation of motion for $\phi$ is 
\begin{eqnarray}
\dalm\phi=0.\label{kg}
\end{eqnarray}
We see from the basic Eqs.~(\ref{beq}) and (\ref{kg}) that if the scalar field in one solution with $\epsilon=1$ is purely imaginary, it can be interpreted as a solution with a ghost scalar field ($\epsilon=-1$).

In this paper, we consider the spherically symmetric spacetime $({\ma M}^4, g_{\mu \nu })$ which is a warped product of a 
two-dimensional constant curvature space $(S^2, \gamma _{ij})$ and a two-dimensional orbit spacetime $(M^2, g_{AB})$ under the isometry of $(S^2, \gamma _{ij})$. 
Namely, the line element is given by
\begin{eqnarray}
ds^2=g_{AB}dx^Adx^B+R(x^A)^2d\Omega^2,\label{sol2}
\end{eqnarray}
where $A,B = 0, 1;~i,j = 2, 3$ and $d\Omega^2:=\gamma _{ij}dx^idx^j=d\theta^2+\sin^2\theta d\varphi^2$.
Here $R$ is a scalar on $(M^2, g_{AB})$ with $R=0$ defining its boundary, and $\gamma_{ij}$ is the unit
metric on $(S^2, \gamma _{ij})$ with its sectional curvature $k = 1$. 
The Misner-Sharp mass~\cite{ms1964} is defined by
\begin{eqnarray}
m_{\rm MS}:=\frac{R}{2}(1-R_{,A}R^{,A}). \label{ms}
\end{eqnarray}

Under the assumption that $(M^2, g_{AB})$ is Minkowski, the general homothetic self-similar spherically symmetric solution for a massless scalar field with $\epsilon=1$ is given by 
\begin{eqnarray}
ds^2=-2dudv+(-uv+C_1v^2+C_2u^2)d\Omega^2, \label{roberts1}
\end{eqnarray}
where $C_1$ and $C_2$ are real constants.
For $C_1C_2=1/4$, it is the Minkowski spacetime.
For $C_1C_2 \ne 1/4$, the scalar field is given by 
\begin{eqnarray}
\phi=\phi_0\pm \frac{1}{2\sqrt{\pi}}{\rm arctanh}\biggl(\frac{1-2C_1(v/u)}{\sqrt{1-4C_1C_2}}\biggl) \label{roberts2}
\end{eqnarray}
for $C_1 \ne 0$ and
\begin{eqnarray}
\phi=\phi_0\pm \frac{1}{4\sqrt{\pi}}\ln\biggl|\frac{v}{u}-C_2\biggl| \label{g-hayward2}
\end{eqnarray}
for $C_1=0$, where the value of the constant $\phi_0$ is meaningless.
The Misner-Sharp mass (\ref{ms}) is given by
\begin{eqnarray}
m_{\rm MS}=-\frac{(1-4C_1C_2)uv}{4\sqrt{-uv+C_1v^2+C_2u^2}}. \label{roberts-mass}
\end{eqnarray}
The Kretschmann invariant $K$ is given by
\begin{eqnarray}
K&:=&{\ma R}_{\mu\nu\rho\sigma}{\ma R}^{\mu\nu\rho\sigma} \nonumber \\
&=&\frac{3u^2v^2(1-4C_1C_2)^2}{(-uv+C_1v^2+C_2u^2)^4}. \label{k-invariant}
\end{eqnarray}

The expression (\ref{roberts2}) is convenient to understand the codomain of $\phi$ because ${\rm arctanh}(w)$ is real, complex, and purely imaginary for $0 \le ({\rm Re}~w)^2<1$ with ${\rm Im}~w=0$, $({\rm Re}~w)^2>1$ with ${\rm Im}~w=0$, and ${\rm Re}~w=0$, respectively.
The scalar field is real for 
\begin{eqnarray}
0 \le \frac{[1-2C_1(v/u)]^2}{1-4C_1C_2}<1 \label{cond1}
\end{eqnarray}
in the case of $C_1C_2<1/4$ otherwise $\phi$ becomes complex.
We write the condition (\ref{cond1}) in terms of the square of the areal radius $R^2=-uv+C_1v^2+C_2u^2$ as
\begin{eqnarray}
0 \le \frac{4C_1R^2}{u^2(1-4C_1C_2)}+1<1.
\end{eqnarray}
Therefore, under $C_1C_2<1/4$, the scalar field is real and complex for $C_1 \le 0$ and $C_1>0$, respectively.
In the case of $C_1C_2>1/4$, on the other hand, Eq.~(\ref{roberts2}) with $\phi_0=0$ can be rewritten as
\begin{eqnarray}
\phi=\pm \frac{i}{2\sqrt{\pi}}\arctan\biggl(\frac{1-2C_1(v/u)}{\sqrt{4C_1C_2-1}}\biggl), \label{roberts-c}
\end{eqnarray}
where $i^2:=-1$.
Therefore, the scalar field is purely imaginary corresponding to a ghost scalar field for $C_1C_2>1/4$.

Here we must give some comments on the history of this solution.
The solution with $C_1 \ne 0$ was obtained by Roberts in 1989~\cite{roberts1989}.
Unfortunately, the metric in the double null coordinates was erroneously written and the correct form was found later by several authors~\cite{sussman1991,ont1994,brady1994,burko1997}.
In this case, we can set $|C_1|=1$ without loss of generality by the coordinate transformations ${\bar v}:=\sqrt{|C_1|}v$ and ${\bar u}:=u/\sqrt{|C_1|}$, so it is a one-parameter family of solutions.
On the other hand, the solution with $C_1=0$ was obtained by Brady in 1994~\cite{brady1994}. (See also~\cite{hayward2000,ch2001}.)
In fact, we can show that the metric of the solution found by Gutman and Bespal'ko for a stiff fluid in 1967~\cite{gb1967} covers half of the spacetime (\ref{roberts1}). 
(See Appendix A.)
This is because a massless scalar field is equivalent to a stiff fluid if the gradient of the scalar field is timelike~\cite{madsen1988}. 
Keeping in mind the history, we call this solution the {\it Roberts solution} in the present paper.


\section{An analytic model of wormhole formation}
In this section, we construct a simple analytic model of wormhole formation by gluing the Roberts and Minkowski spacetimes in a regular manner.
We focus on the case of $C_1C_2>1/4$ corresponding to a ghost scalar field, which is required for this construction.
The properties of the Roberts solution in the case of $C_1C_2 \le 1/4$ is reviewed in Appendix B.

In the case of $C_1C_2>1/4$, $C_1>0$ and $C_2>0$ are required for the areal radius to be nonnegative.
The areal radius becomes zero only at $u=v=0$.
Thus, it is seen in Eq.~(\ref{k-invariant}) that only $u=v=0$ may be a curvature singularity.
We also see in Eq.~(\ref{roberts-mass}) that the region with $uv>(<)0$ has positive (negative) mass.
On a null hypersurface of $u=0$ or $v=0$, the Kretschmann invariant and the quasilocal mass are zero and the derivative of the scalar field becomes null.

The trapped region is given by $R<2m_{\rm MS}$.
Since $m_{\rm MS} \le 0$ is satisfied for $uv \le 0$, the trapped region is located in the region of $uv>0$.
The trapping horizon~\cite{hayward1994} defined by $R=2m_{\rm MS}$ is given by
\begin{eqnarray}
v=-\frac{(1+4C_1C_2)\pm \sqrt{1+8C_1C_2}}{4C_1}u \label{th}
\end{eqnarray}
for $C_1 \ne 0$, while $u=0$ and $v=2C_2 u$ for $C_1=0$.
Thus, there are two timelike trapping horizons (\ref{th}) for $C_1C_2>1/4$ with $C_1, C_2>0$ in the region of $uv>0$.

\begin{figure}[htbp]
\begin{center}
\includegraphics[width=0.7\linewidth]{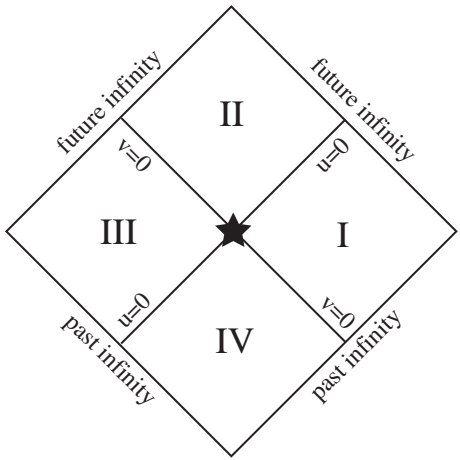}
\caption{\label{Penrose4}
The Penrose diagram of the Roberts solution (\ref{roberts1}) for $C_1C_2>1/4$ with $C_1>0$ and $C_2>0$.
A star corresponds to an instantaneous curvature singularity at $u=v=0$.
The quasilocal mass is positive (negative) in the regions II and IV (I and III).
The trapping horizons (\ref{th}) are located in the regions II and IV. 
}
\end{center}
\end{figure}

It is easy to know the global structure of the Roberts solution because $(M^2, g_{AB})$ is the Minkowski spacetime.
$u$ and $v$ are affine parameters along radial null geodesics, and then null infinities are represented by $u \to \pm\infty$ or $v \to \pm\infty$.
The Penrose diagram of the Roberts spacetime for $C_1C_2>1/4$ with $C_1, C_2>0$ is given in Fig.~\ref{Penrose4}.
This spacetime represents a dynamical wormhole. 
(Several (quasi-)local definitions of a dynamical wormhole have been independently proposed so far~\cite{hayward1999,hv1998,mhc2009}.)

Now we show that the Roberts spacetime can be attached to the Minkowski spacetime at $u=0$ or $v=0$ in a regular manner, i.e., without a massive thin shell on the hypersurface. 
(See~\cite{bi1991,Poisson} for the matching condition on a null hypersurface.)
We consider a null hypersurface $u=0$ as a matching surface, which we call $\Sigma$.
(The argument is similar for $v=0$.)
The induced metric $h_{ab}$ on $\Sigma$ is given by
\begin{eqnarray}
ds_{\Sigma}^2=h_{ab}dy^a dy^b:=C_1v^2d\Omega^2,
\end{eqnarray}
where $y^a=(v,\theta,\varphi)$ is a set of coordinates on $\Sigma$.
The basis vectors of $\Sigma$ defined by $e^\mu_a := \partial x^\mu/\partial y^a$ are given by
\begin{align}
e^\mu_v\frac{\partial}{\partial x^\mu}&=\frac{\partial}{\partial v},\\
e^\mu_i\frac{\partial}{\partial x^\mu}&=\delta^\mu_{~~i}\frac{\partial}{\partial x^i}.
\end{align}
The basis is completed by $N_\mu dx^\mu=-dv$ satisfying $N_\mu e^\mu_v=-1$ and $N_\mu e^\mu_i=0$ on $\Sigma$.
The only nonvanishing component of the transverse curvature $C_{ab}:=N_{\mu;\nu} e^\mu_{a} e^\nu_b$ of $\Sigma$ is 
\begin{equation}
C_{ij}=-\frac12 v \gamma_{ij}.
\end{equation}

The regular attachment on $\Sigma$ requires the continuity of $h_{ab}$ and $C_{ab}$ on both side of $\Sigma$.
Since there is no $C_2$ in the expressions of $h_{ab}$ and $C_{ab}$, two Roberts spacetimes with the same nonzero $C_1$ but different $C_2$ can be attached in a regular manner at $u=0$.
Thus, as a special case, the Roberts spacetime (\ref{roberts1}) with $C_1={\bar C}_1(\ne 0)$ and $C_2={\bar C}_2$ can be attached to the past Minkowski spacetime at $u=0$ in a regular manner, of which metric is given by Eq.~(\ref{roberts1}) with $C_1={\bar C}_1$ and $C_2=1/(4{\bar C}_1) \ne {\bar C}_2$.
Similarly, it is shown that two Roberts spacetimes with the same nonzero $C_2$ but different $C_1$ can be attached in a regular manner at $v=0$.

\begin{figure}[htbp]
\begin{center}
\includegraphics[width=1.0\linewidth]{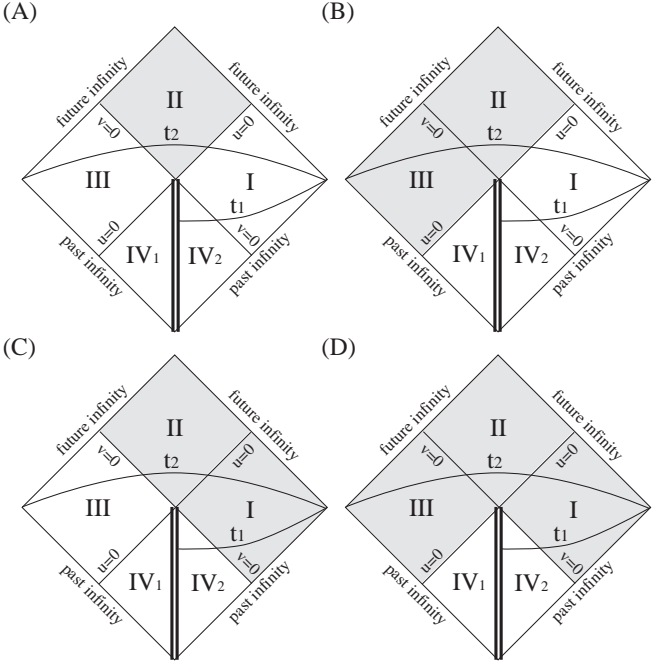}
\caption{\label{Penrose2}
The Penrose diagrams representing wormhole formation from the initial data with a regular center.
The Roberts spacetime with $C_1C_2>1/4$, $C_1>0$, and $C_2>0$ (the shadowed region) is attached to the past Minkowski spacetimes at (A) $u=0$ with $v>0$ and $v=0$ with $u>0$, (B) $u=0$, (C) $v=0$ with $u>0$, and (D) $u=0$ with $v<0$ and $v=0$ with $u<0$.
A thick line corresponds to a symmetric center in a Minkowski spacetime.
$t_1$ represents a spacelike hypersurface with a regular symmetric center, while $t_2$ represents a spacelike hypersurface with distinct spacelike infinities without a regular center.
}
\end{center}
\end{figure}

By gluing the Roberts spacetime with $4C_1C_2>1$, $C_1,C_2>0$ and the Minkowski spacetime(s) at $u=0$ and/or $v=0$ in a regular manner, we can construct spacetimes representing wormhole formation from the initial data with a regular center.
The Penrose diagrams for these spacetimes are given in Fig.~\ref{Penrose2}.

The attachment of the Roberts spacetime to the Minkowski spacetime in the case of the ghost scalar field has been mentioned in~\cite{fj2004} without detailed calculations.
It is claimed there that the instantaneous singularity $u=v=0$ in the Roberts spacetime can be removed by gluing the Minkowski spacetime at $u=0$ or $v=0$.
Obviously, the curvature invariants do not blow up if an observer approaches there from the Minkowski region, however, they certainly blow up along some causal geodesics emanating from $u=v=0$ in the Roberts region.
As a result, there is still a naked singularity at $u=v=0$ in the resulting spacetime.
The details will be presented in the next section.


\section{Properties of the instantaneous singularity}
In the last section, we constructed a spacetime representing the wormhole formation.
One problem in this spacetime is an instantaneous curvature singularity at $u=v=0$ which appears at the moment of the wormhole formation.
In this section, we show that it is a naked but weak singularity.

\subsection{Nakedness}
First we show that both radial and nonradial causal geodesics emanate from $u=v=0$, i.e., it is certainly a naked singularity.
The Lagrangian to give the geodesic equations is 
\begin{align}
L=&\frac12g_{\mu\nu}{\dot x}^\mu{\dot x}^\nu \nonumber \\
=&-{\dot u}{\dot v}+\frac12(-uv+C_1v^2+C_2u^2)({\dot \theta}^2+\sin^2\theta{\dot \varphi}^2).
\end{align}  
where a dot denotes the derivative with respect to the affine parameter $\lambda$ along a geodesic.
Because of spherical symmetry, we can set $\theta \equiv \pi/2$ without loss of generality.
The metric (\ref{roberts1}) is independent of $\varphi$, so that from the Lagrange equation 
\begin{align}
0=\frac{\partial}{\partial \lambda}\frac{\partial L}{\partial {\dot x}^\mu}-\frac{\partial L}{\partial {x}^\mu},\label{lag}
\end{align}  
we obtain a conserved quantity along a geodesic as
\begin{align}
\Phi&:= \frac{\partial L}{\partial {\dot \varphi}} \nonumber \\
&=(-uv+C_1v^2+C_2u^2){\dot \varphi}. \label{angC}
\end{align}  
Then, the geodesic equations (\ref{lag}) are written as
\begin{align}
0&={\ddot v}+\frac{(-v+2C_2u)\Phi^2}{2(-uv+C_1v^2+C_2u^2)^2},\\
0&={\ddot u}+\frac{(-u+2C_1v)\Phi^2}{2(-uv+C_1v^2+C_2u^2)^2}.
\end{align}  

The tangent vector of a nonspacelike geodesic $k^\mu:={\dot x^\mu}$ satisfies
\begin{align}
k^\mu k_\mu=\varepsilon,
\end{align}  
where $\varepsilon$ is $0$ and $-1$ for null and timelike geodesics, respectively.
This equation is written as
\begin{align}
\varepsilon=-2{\dot u}{\dot v}+(-uv+C_1v^2+C_2u^2)^{-1}\Phi^2. 
\end{align}  

The Roberts spacetime admits a homothetic Killing vector $\xi^\mu(\partial/\partial x^\mu)=u(\partial/\partial u) +v(\partial/\partial v)$ satisfying 
\begin{align}
{\cal L}_{\xi}g_{\mu\nu}:=\xi_{\mu;\nu}+\xi_{\nu;\mu}=2g_{\mu\nu}.\label{homo}
\end{align}  
Then, we obtain
\begin{align}
\frac{d}{d\lambda}(\xi^\mu k_\mu)&=:(\xi^\mu k_\mu)_{;\nu}k^\nu \nonumber \\
&=\xi_{\mu;\nu}k^\mu k^\nu+\xi^\mu k_{\mu;\nu}k^\nu \nonumber \\ 
&=\xi_{\mu;\nu}k^\mu k^\nu \nonumber \\
&=g_{\mu\nu}k^\mu k^\nu \nonumber \\
&=\varepsilon,
\end{align}  
where we used the fact that $k^\mu$ is tangent to a geodesic and Eq.~(\ref{homo}).
Hence we obtain $\xi^\mu k_\mu=D_0+\varepsilon \lambda$, or equivalently 
\begin{align}
-u{\dot v}-v{\dot u}=D_0+\varepsilon \lambda, \label{eq-1}
\end{align}  
where $D_0$ is a constant.
Equation~(\ref{eq-1}) is integrated to give
\begin{align}
-uv=D_1+D_0\lambda+\frac12 \varepsilon \lambda^2,
\end{align}  
where $D_1$ is a constant.

We are now interested in the geodesics emanating from $u=v=0$.
Without loss of generality, we can set $\lambda$ such that $\lambda=0$ corresponds to $u=v=0$.
Thus, we consider the case with $D_1=0$.
Now the geodesic equations reduce to
\begin{align}
\varepsilon&=-2{\dot u}{\dot v}+(-uv+C_1v^2+C_2u^2)^{-1}\Phi^2, \label{key2} \\
-uv&=D_0\lambda+\frac12 \varepsilon \lambda^2. \label{key1}
\end{align}  
We obtain the master equation for $u(\lambda)$ from above equations  as
\begin{align}
\frac{du}{d\lambda}&=\frac{Au(\varepsilon \lambda+D_0)\pm u\sqrt{A[D_0^2A+4\lambda\Phi^2u^2(\varepsilon \lambda+2D_0)]}}{\lambda A(\varepsilon \lambda+2D_0)},\label{master} \\
A&:=4C_2 u^4+2\lambda u^2(\varepsilon \lambda+2D_0)+C_1\lambda^2(\varepsilon \lambda+2D_0)^2.
\end{align}  

First we consider radial geodesics ($\Phi=0$).
For the null geodesics ($\varepsilon=0$), the solutions of Eqs.~(\ref{key2}) and (\ref{key1}) passing through $u=v=0$ are $u=0$ or $v=0$.
Along these radial null geodesics, the Kretschmann invariant and the quasilocal mass are identically zero. 
For the timelike geodesics ($\varepsilon=-1$), the general solution of Eq.~(\ref{master}) is given by $u=u_0\lambda$ and $u=u_0(2D_0-\lambda)$, where $u_0$ is a nonzero constant.
These two coincide for $D_0=0$ and the latter does not pass through $u=0$ for $D_0 \ne 0$.
Finally, the solution passing through $u=v=0$ is given by 
\begin{align}
u=u_0\lambda, \quad v=\frac{1}{2u_0}\lambda, \label{t-radial}
\end{align}  
which corresponds to $D_0=0$ in Eq.~(\ref{key1}).
The Kretschmann invariant along these radial timelike geodesics is given by 
\begin{eqnarray}
K=\frac{192u_0^8(1-4C_1C_2)^2}{(-2u_0^2+C_1+4C_2u_0^4)^4\lambda^4},
\end{eqnarray}
which diverges at $\lambda=0$, i.e., $u=v=0$.
Along these radial timelike geodesics, the quasilocal mass is given by
\begin{eqnarray}
m_{\rm MS}=-\frac{(1-4C_1C_2)u_0\lambda}{4\sqrt{-2u_0^2+C_1+4C_2u_0^4}},
\end{eqnarray}
which converges to zero at $\lambda=0$.

For the nonradial geodesics ($\Phi \ne 0$), there is a solution of Eqs.~(\ref{key2}) and (\ref{key1}) passing through $u=v=0$, which is given by
\begin{widetext}
\begin{align}
u^2&=\frac{-\lambda(2D_0+\varepsilon \lambda)[(D_0^2+2\Phi^2)\pm\sqrt{(D_0^2+2\Phi^2)^2-4D_0^4C_1C_2}]}{4D_0^2C_2}, \label{u} \\
v^2&=\frac{-\lambda(2D_0+\varepsilon \lambda)[(D_0^2+2\Phi^2)\mp\sqrt{(D_0^2+2\Phi^2)^2-4D_0^4C_1C_2}]}{4D_0^2C_1}. \label{v}
\end{align}  
\end{widetext}
Under $4C_1C_2-1>0$ and $C_1, C_2>0$, the conditions for $u^2$ and $v^2$ to be real and positive are $D_0<0$ and 
\begin{align}
\frac14<C_1C_2 \le \frac{(D_0^2+2\Phi^2)^2}{4D_0^4}.\label{non-rad}
\end{align}  
Because $D_0$ and $\Phi$ are independent parameters which characterize a geodesic, the right-hand side of Eq.~(\ref{non-rad}) varies from $1/4$ to infinity.
Hence, for any values of $C_1$ and $C_2$ satisfying $C_1C_2>1/4$ and $C_1, C_2>0$, there are nonradial causal geodesics passing through $u=v=0$.

Along these nonradial geodesics, the Kretschmann invariant diverges around $u=v=0$ as
\begin{align}
\lim_{\lambda \to 0}K &\simeq \frac{3u_0v_0(1-4C_1C_2)^2}{(-u_0v_0+C_1v_0^2+C_2u_0^2)^4\lambda^2}, \label{K-non}\\
u_0^2&:=\frac{-(D_0^2+2\Phi^2)\mp\sqrt{(D_0^2+2\Phi^2)^2-4D_0^4C_1C_2}}{2D_0C_2}, \label{u0} \\
v_0^2&:=\frac{-(D_0^2+2\Phi^2)\pm\sqrt{(D_0^2+2\Phi^2)^2-4D_0^4C_1C_2}}{2D_0C_1}, \label{v0}
\end{align}
while the quasilocal mass converges to zero at $u=v=0$ as
\begin{eqnarray}
\lim_{\lambda \to 0}m_{\rm MS} \simeq-\frac{(1-4C_1C_2)u_0v_0\lambda^{1/2}}{4\sqrt{-u_0v_0+C_1v_0^2+C_2u_0^2}}. \label{m-non}
\end{eqnarray}
We note that Eqs.~(\ref{K-non}) and (\ref{m-non}) are exact expressions for nonradial null geodesics.

\subsection{Strength}
Next we consider the strength of the singularity at $u=v=0$.
As definitions of the strength of singularities, the strong curvature condition (SCC)~\cite{Tipler} and the limiting focusing condition (LFC)~\cite{Krolak} were proposed by Tipler and Kr\'{o}lak, respectively.
We consider a geodesic (N), affinely parametrized by $\lambda$, with the tangent vector $k^\mu$, terminating at or emanating from a singularity, where $\lambda=0$.
SCC and LFC imply that N is emanating from or terminating in the Tipler's strong and the Kr\'{o}lak's strong curvature singularities, respectively~\cite{ClarkeKrolak}.
The physical content of the Tipler strong is that the volume element of physical objects (constructed by the Jacobi fields along N) converges to zero at the singularity.
On the other hand, the physical content of the Kr\'{o}lak strong is that the expansion along N diverges at the singularity, but still the volume element remains finite.
(See also~\cite{Joshi,Clarketext} for the textbook.)

The necessary conditions for SCC and LFC are available~\cite{ClarkeKrolak}.
Let $E_{(I)}^\mu (I=1,2,3,4)$ a parallelly propagating frame along N satisfying $E^\mu_{(1)}E_{(1)\mu}=E^\mu_{(2)}E_{(2)\mu}=E^\mu_{(3)}E_{(3)\mu}=-E^\mu_{(4)}E_{(4)\mu}=1$ if N is timelike and $E^\mu_{(1)}E_{(1)\mu}=E^\mu_{(2)}E_{(2)\mu}=-E^\mu_{(3)}E_{(4)\mu}=-E^\mu_{(4)}E_{(3)\mu}=1$ if N is null.
All other products vanish and $E_{(4)\mu}:=k^\mu$.
If SCC is satisfied along N, then $\lim_{\lambda \to 0}\lambda^2{\cal R}^{(I)}_{~~~(4)(J)(4)}$ does not converge for $I,J=1,2,3$ and $I,J=1,2$ in the cases where N is timelike and null, respectively, where ${\cal R}^{(I)}_{~~~(J)(K)(L)}$ is the Riemann tensor in the parallelly propagating frame.
If LFC is satisfied along N, then $\lim_{\lambda \to 0}\lambda {\cal R}^{(I)}_{~~~(4)(J)(4)}$ does not converge for $I,J=1,2,3$ and $I,J=1,2$ in the cases where N is timelike and null, respectively.

We show that the singularity at $u=v=0$ in the Roberts spacetime is weak in the senses of both Tipler and Kr\'{o}lak for radial causal geodesics.
For the radial causal geodesics, of which tangent vector has the form of $k^\mu=(k^u,k^v,0,0)=:E^\mu_{(4)}$, where $k^u k^v=0$ is satisfied for null geodesics, the angler bases $E^\mu_{(I)}$ ($I=1,2$) are given as
\begin{align}
E^\mu_{(1)}\frac{\partial}{\partial x^\mu}&:=\frac{1}{R}\frac{\partial}{\partial \theta},\\
E^\mu_{(2)}\frac{\partial}{\partial x^\mu}&:=\frac{1}{R\sin\theta}\frac{\partial}{\partial \varphi},
\end{align}  
which satisfy $E^\mu_{(I)}E_{(J)\mu}=\delta_{IJ}$.
The only nonzero component of ${\cal R}^{\mu}_{~~(4)\nu(4)}:={\cal R}^{\mu}_{~\rho\nu\sigma}k^\rho k^\sigma$ is
\begin{align}
{\cal R}^{i}_{~~(4)j(4)}=-\frac{(4C_1C_2-1)(vk^u-uk^v)^2}{4(-uv+C_1v^2+C_2u^2)^2}\delta^i_{~~j}.
\end{align}  
Thus, the only nonzero component of ${\cal R}^{(I)}_{~~~(4)(J)(4)}$ is 
\begin{align}
{\cal R}^{(1)}_{~~(4)(1)(4)}={\cal R}^{(2)}_{~~(4)(2)(4)}=-\frac{(4C_1C_2-1)(vk^u-uk^v)^2}{4(-uv+C_1v^2+C_2u^2)^2}.
\end{align}  
These quantities are identically zero both for radial null geodesics ($u=0$ or $v=0$) and radial timelike geodesics (\ref{t-radial}).
Hence, it is concluded that the singularity at $u=v=0$ is weak in the senses of both Tipler and Kr\'{o}lak for radial causal geodesics.

For nonradial geodesics, it seems to be cumbersome to examine ${\cal R}^{(I)}_{~~~(4)(J)(4)}$.
Instead, we here consider the behavior of ${\cal R}_{(4)(4)}:={\cal R}_{\mu\nu}k^\mu k^\nu$, which is used in the sufficient conditions for SCC and LFC.
SCC is satisfied if $\lim_{\lambda \to 0}\lambda^2{\cal R}_{(4)(4)}>0$ and LFC is satisfied if $\lim_{\lambda \to 0}\lambda {\cal R}_{(4)(4)}>0$~\cite{ClarkeKrolak,Clarketext}. 
The nonzero components of the Ricci tensor of the Roberts spacetime are given by
\begin{align}
{\cal R}_{uu}&=-\frac{v^2(4C_1C_2-1)}{2(-uv+C_1v^2+C_2u^2)^2},\\
{\cal R}_{uv}&=\frac{uv(4C_1C_2-1)}{2(-uv+C_1v^2+C_2u^2)^2},\\
{\cal R}_{vv}&=-\frac{u^2(4C_1C_2-1)}{2(-uv+C_1v^2+C_2u^2)^2}.
\end{align}  
Finally, for nonradial causal geodesics $k^\mu=({\dot u},{\dot v},0,{\dot \varphi})$, where ${\dot u}$, ${\dot v}$, and ${\dot \varphi}$ are obtained from (\ref{u}), (\ref{v}), and (\ref{angC}), respectively, we obtain ${\cal R}_{(4)(4)}\equiv 0$, which immediately implies $\lambda^2 {\cal R}_{(4)(4)} \equiv 0$ and $\lambda {\cal R}_{(4)(4)} \equiv 0$ along the geodesics.
Although this result does not directly mean that neither SCC nor LFC is satisfied, it would suggest that the singularity at $u=v=0$ is weak also along nonradial causal geodesics.


\section{Summary and discussions}
In this paper, we constructed an explicit and simple model of wormhole formation from the initial data with a regular center.
The spacetime represents the wormhole formation with a massless ghost scalar field.
In this construction, the matter region represented by the Roberts spacetime is attached to the past Minkowski spacetimes at null hypersurfaces in a regular manner.

This construction has been mentioned in~\cite{fj2004} without a detailed analysis.
Actually, unlike the authors' claim in~\cite{fj2004}, a naked curvature singularity appears at the moment of the wormhole formation. 
However, we showed that it is only instantaneous and weak in the senses of both Tipler and Kr\'{o}lak.
This class of weak singularities could be harmless because it would be dealt with in some distributional sense.

In this context, Hayward and Koyama constructed an analytic model representing the wormhole ``formation'' from the Schwarzschild black hole~\cite{kh2004}.
Although their model does not contain a singularity at the moment of ``formation'', it does not represent the wormhole formation from the initial data with a regular center, i.e., there is no topology change in their model.
While they defined a wormhole throat by a class of trapping horizons~\cite{hayward1999,hv1998}, there is a wormhole throat on any spacelike hypersurface in their model.
(See~\cite{mhc2009} for the discussions of the (quasi-)local definition of a wormhole throat on a spacelike hypersurface.)

In the spatially compact spacetime, the wormhole formation necessarily requires the appearance of singularities or closed timelike curves.
This is a consequence of the result by Geroch about the topology change of spacetimes~\cite{geroch1967}.
Hence, under the suitable chronology condition, the singularity formation is inevitable.
In this paper, on the other hand, the wormhole formation in the spatially noncompact spacetime is considered.
The singularity formation would be also inevitable for the wormhole formation in this case, however, as far as the author knows, this is an open question.

The solution presented in this paper will be a simple analytic model to study the formation of a wormhole.
In this context, the stability of the wormhole formation is an important future work.
While the stability of the Roberts solution was studied in the case of the positive kinetic term of the scalar field~\cite{frolov1997}, it is still an open question for the ghost scalar field.

\section*{Acknowledgments}
The author would like to thank T~Harada, K.~Nakao, J.~Oliva, J.M.M.~Senovilla, and R.~Vera for discussions and comments. 
The author was supported by Fondecyt Grant No. 1071125.
The Centro de Estudios Cient\'{\i}ficos (CECS) is funded by the Chilean
Government through the Millennium Science Initiative and the Centers of
Excellence Base Financing Program of Conicyt. CECS is also supported by a
group of private companies which at present includes Antofagasta Minerals,
Arauco, Empresas CMPC, Indura, Naviera Ultragas, and Telef\'{o}nica del Sur.
The author is also grateful to Universidad del Pais Vasco, Universidade de Santiago de Compostela, and Instituto Superior T{\'e}cnico for their hospitality during his visit.

\appendix
\section{The relation between the Gutman-Bespal'ko and the Roberts solutions}
In 1967, Gutman and Bespal'ko obtained a spherically symmetric solution for a stiff fluid, i.e., a perfect fluid with an equation of state $p=\mu$, where $p$ and $\mu$ are the pressure and energy density, respectively~\cite{gb1967}.
(See also~\cite{wesson1978,krasinski} for the generalized solution.)
The energy-momentum tensor for a perfect fluid is given by 
\begin{equation}
T_{\mu\nu}=pg_{\mu\nu}+(p+\mu)u_\mu u_\nu, \label{pf}
\end{equation}
where $u^\mu$ is the four-velocity of the fluid element.
The Gutman-Bespal'ko solution is given in the comoving coordinates as
\begin{eqnarray}
ds^2&=&-z^2dt^2+dz^2 \nonumber \\
&&~~~~~~+\frac12 z^2(1+C_1e^{2t}+C_2e^{-2t})d\Omega^2, \label{gb} \\
p&=&\mu=\frac{1-4C_1C_2}{8\pi z^2(1+C_1e^{2t}+C_2e^{-2t})^2},\label{sol2}
\end{eqnarray}
where the constants $C_1$ and $C_2$ satisfy $C_1C_2 \le 1/4$ for nonnegative energy density.
In the case of $4C_1C_2=1$, the solution gives the Minkowski solution.
The Gutman-Bespal'ko spacetime admits a homothetic Killing vector $\xi^\mu(\partial/\partial x^\mu)=z(\partial/\partial z)$ satisfying ${\cal  L}_{\xi} g_{\mu\nu} = 2g_{\mu\nu} $.

We show that the Gutman-Bespal'ko spacetime covers half of the Roberts spacetime.
In 1988, Madsen showed the equivalence between a massless scalar field and a stiff fluid~\cite{madsen1988}.
This is easily generalized to the ghost case as shown below.
The energy-momentum tensor for a stiff fluid is 
\begin{equation}
T_{\mu\nu}=\mu(2u_{\mu}u_{\nu}+g_{\mu\nu}).\label{s-fluid}
\end{equation}
If $u^{\mu}$ is vorticity free, which is satisfied in the spherically symmetric spacetime, one can show that this matter field is equivalent to a massless scalar field $\phi$, of which gradient $\phi_{,\mu}$ is timelike. 
The corresponding energy density and 4-velocity are given by
\begin{eqnarray}
\mu &=& -\frac{1}{2}\epsilon \phi^{,\mu}\phi_{,\mu},\\ 
u_{\mu}&=&\pm \frac{\phi_{,\mu}}{\sqrt{-\phi_{,\nu}\phi^{,\nu}}},\label{eq:epsilon}
\end{eqnarray}
with which Eq.~(\ref{s-fluid}) coincides with Eq.~(\ref{eq:stress-energy_tensor_of_scalar_field}), where the sign in (\ref{eq:epsilon}) is chosen so that $u^{\mu}$ is future-directed.

For the Gutman-Bespal'ko solution, the corresponding scalar field with $\epsilon=1$ is
\begin{eqnarray}
\phi=\phi_0 \pm \frac{1}{2\sqrt{\pi}}{\rm arctanh}\biggl(\frac{1+2C_1e^{2t}}{\sqrt{1-4C_1C_2}}\biggl)
\end{eqnarray}
for $C_1 \ne 0$ and 
\begin{eqnarray}
\phi=\phi_0\pm \frac{1}{4\sqrt{\pi}}\ln\biggl|-e^{2t}-C_2\biggl|
\end{eqnarray}
for $C_1=0$.

In the Gutman-Bespal'ko spacetime, the metric on $(M^2,g_{AB})$ is Minkowski in the Rindler coordinates, while it is also Minkowski but in the double null coordinates in the Roberts solution.
By the coordinate transformations
\begin{eqnarray}
t={\rm arctanh}\biggl(\frac{T}{X}\biggl), ~~~z=\pm \sqrt{X^2-T^2},
\end{eqnarray}
of which inverse transformations are
\begin{eqnarray}
T=z\sinh t, ~~~X=z\cosh t,
\end{eqnarray}
the two-dimensional Rindler metric $ds_2^2=-z^2dt^2+dz^2$ is transformed into $ds_2^2=-dT^2+dX^2$.
Adopting the null coordinates such as
\begin{eqnarray}
u=\frac{T-X}{\sqrt{2}}, ~~~v=\frac{T+X}{\sqrt{2}},
\end{eqnarray}
we obtain $ds_2^2=-2dudv$.

Indeed, by the direct transformations
\begin{eqnarray}
&&u=\pm\frac{1}{\sqrt{2}}ze^{-t}, ~~~v=\mp\frac{1}{\sqrt{2}}ze^{t}, \label{trans-1a}
\end{eqnarray}
of which inverse is
\begin{eqnarray}
&&t={\rm arctanh}\biggl(\frac{v+u}{v-u}\biggl), ~~~z=\pm \sqrt{-2uv}, \label{trans-1b}
\end{eqnarray}
the Roberts metric (\ref{roberts1}) is transformed to the Gutman-Bespal'ko metric (\ref{gb}).
Therefore, we may call the Gutman-Bespal'ko metric (\ref{gb}) the {\it Rindler chart} of the Roberts metric. 
Because of $uv<0$, the Rindler chart covers only half of the Roberts spacetime.
(The regions I and III in Figs.~\ref{Penrose4} and \ref{Penrose}.)

On the other hand, by the coordinate transformations
\begin{eqnarray}
T={\tilde t}\cosh {\tilde z}, ~~~X={\tilde t}\sinh {\tilde z},
\end{eqnarray}
of which inverse is
\begin{eqnarray}
{\tilde t}=\pm \sqrt{T^2-X^2},~~{\tilde z}={\rm arctanh}\biggl(\frac{X}{T}\biggl),
\end{eqnarray}
the two-dimensional Minkowski spacetime $ds_2^2=-dT^2+dX^2$ is transformed to the Milne form $ds_2^2=-d{\tilde t}^2+{\tilde t}^2d{\tilde z}^2$.
Thus, by the direct transformations
\begin{eqnarray}
u=\pm\frac{1}{\sqrt{2}}{\tilde t}e^{-{\tilde z}}, ~~~v=\pm\frac{1}{\sqrt{2}}{\tilde t}e^{\tilde z}, \label{trans-2a}
\end{eqnarray}
of which inverse is
\begin{eqnarray}
{\tilde t}=\pm\sqrt{2uv},~~~{\tilde z}={\rm arctanh}\biggl(\frac{v-u}{v+u}\biggl), \label{trans-2b}
\end{eqnarray}
the Roberts solution is transformed to
\begin{eqnarray}
ds^2&=&-d{\tilde t}^2+{\tilde t}^2d{\tilde z}^2 \nonumber \\
&&~~+\frac12{\tilde t}^2(-1+C_1e^{2{\tilde z}}+C_2e^{-2{\tilde z}})d\Omega^2,\label{roberts-milne}\\
\phi&=&\phi_0 \pm \frac{1}{2\sqrt{\pi}}{\rm arctanh}\biggl(\frac{1-2C_1e^{2{\tilde z}}}{\sqrt{1-4C_1C_2}}\biggl)
\end{eqnarray}
for $C_1 \ne 0$. 
For $C_1=0$, the scalar field is transformed to 
\begin{eqnarray}
\phi=\phi_0\pm \frac{1}{4\sqrt{\pi}}\ln\biggl|e^{2{\tilde z}}-C_2\biggl|.
\end{eqnarray}
We may call this metric the {\it Milne chart} of the Roberts metric. 
Because of $uv>0$, the Milne chart covers the regions II and IV in Figs.~\ref{Penrose4} and \ref{Penrose}.
This spacetime admits a homothetic Killing vector $\xi^\mu(\partial/\partial x^\mu)=t(\partial/\partial t)$ satisfying ${\cal  L}_{\xi} g_{\mu\nu} = 2g_{\mu\nu} $.

\section{The Roberts solution for $C_1C_2 \le 1/4$}
\begin{figure}[htbp]
\begin{center}
\includegraphics[width=1.0\linewidth]{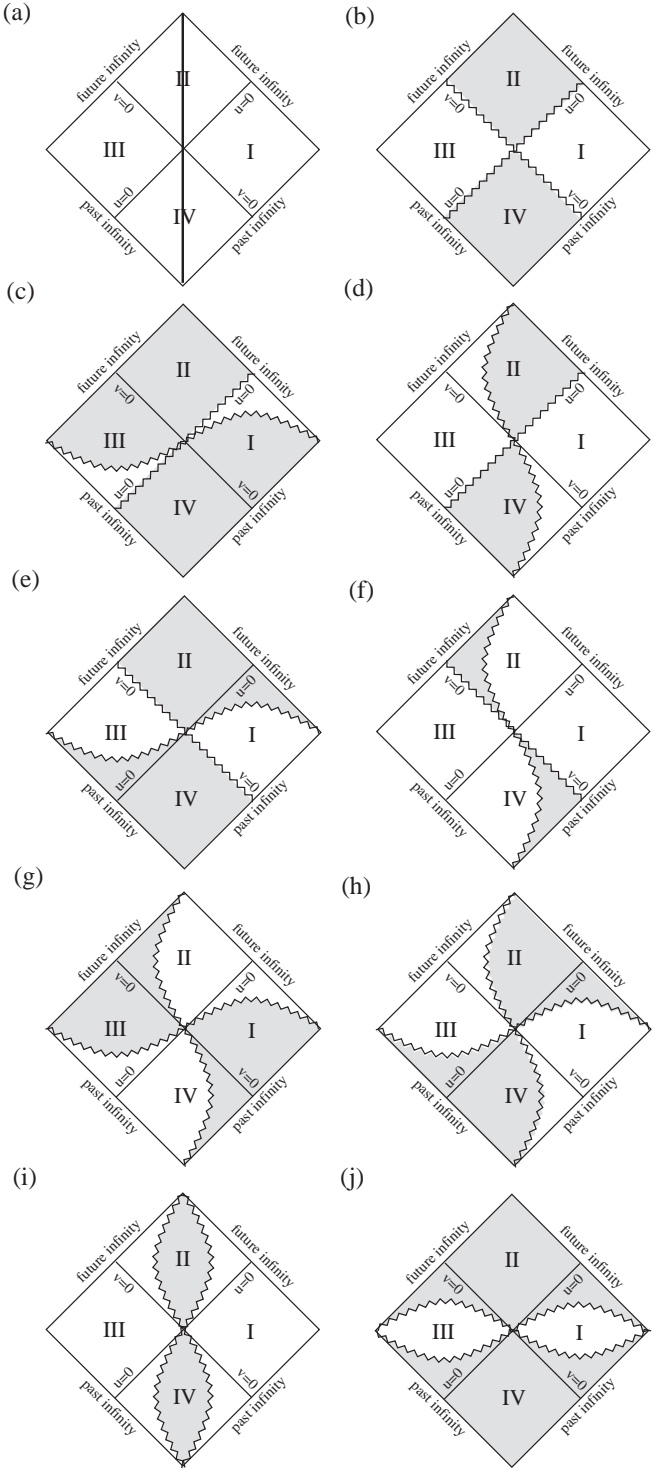}
\caption{\label{Penrose}
The Penrose diagrams of the Roberts solution for (a) $C_1C_2=1/4$ (Minkowski), (b) $C_1=C_2=0$, (c) $C_1=0$ and $C_2<0$, (d) $C_1=0$ and $C_2>0$, (e) $C_2=0$ and $C_1<0$, (f) $C_2=0$ and $C_1>0$, (g) $C_1>0$ and $C_2<0$, (h) $C_1<0$ and $C_2>0$, (i) $0<C_1C_2<1/4$ and $C_1>0$, and (j) $0<C_1C_2<1/4$ and $C_1<0$.
A thick line in (a) corresponds to the symmetric center.
A zigzag curve corresponds to a curvature singularity.
The areal radius is negative in the shadowed region, which is unphysical.
}
\end{center}
\end{figure}

In this appendix, we review the properties of the Roberts spacetime with $C_1C_2 \le 1/4$ corresponding to the positive kinetic term of the scalar field.
First we see in Eq.~(\ref{roberts-mass}) that the region with $uv>0$ has negative mass.
In the case with $C_1 \ne 0$, $C_2 \ne 0$, and $C_1C_2 \ne 1/4$, there are nonnull central curvature singularities located at
\begin{eqnarray}
u=\frac{1\pm\sqrt{1-4C_1C_2}}{2C_1}v.
\end{eqnarray}
If $C_1C_2>0$, both of them are timelike or spacelike, while if $C_1C_2<0$, one is spacelike and the other is timelike.
For $C_1=0$ and $C_2 \ne 0$, there are null and nonnull central curvature singularities at $u=0$ and $u=(1/C_2)v$, respectively.
For $C_2=0$ and $C_1 \ne 0$, there are null and nonnull central curvature singularities at $v=0$ and $v=(1/C_1)u$, respectively.
For $C_1=0$ and $C_2=0$, there are null central curvature singularities at $u=0$ and $v=0$.
The Penrose diagram of the Roberts spacetime for $C_1C_2 \le 1/4$ is given in Fig.~\ref{Penrose}.

As shown in the main text, the Roberts spacetime can be attached to the Minkowski spacetime at a null hypersurface $u=0$ or $v=0$ in a regular manner if that hypersurface is regular.
The resulting spacetime can be a model of the gravitational collapse leading to the naked singularity formation.
This model has been considered in the context of critical phenomena or cosmic censorship~\cite{roberts1989,brady1994,ont1994,brady1995,frolov1997,hayward2000,application}.


\begin{thebibliography}{99}
\bibitem{superluminal}
  M.~Visser, B.~Bassett, and S.~Liberati,
  arXiv:gr-qc/9908023;
  M.~Visser, B.~Bassett, and S.~Liberati,
  Nucl.\ Phys.\ Proc.\ Suppl.\  {\bf 88}, 267 (2000);
\bibitem{visser}
M. Visser, {\it Lorentzian Wormholes: From Einstein to Hawking},
(Springer-Verlag, Berlin, Germany, 1997).
\bibitem{lobo2007}
F.S.N.~Lobo, 
e-Print: arXiv:0710.4474 [gr-qc]. 
\bibitem{mty1988}
M.S.~Morris, K.S.~Thorne, and U.~Yurtsever, 
Phys. Rev. Lett. {\bf 61}, 1446 (1988).
\bibitem{timemachine}
M.~Visser,
Phys. Rev. D{\bf 47}, 554 (1993);
S.W.~Kim and K.S.~Thorne, 
Phys. Rev. D{\bf 43}, 3929 (1991). 
\bibitem{mt1988}
M.S.~Morris and K.S.~Thorne,
Am. J. Phys. {\bf 56}, 395 (1988).
\bibitem{before}
  H.~G.~Ellis,
  J.\ Math.\ Phys.\  {\bf 14}, 104 (1973);
  H.~G.~Ellis,
  Gen.\ Rel.\ Grav.\  {\bf 10}, 105 (1979);
  K.~A.~Bronnikov,
  Acta Phys.\ Polon.\  B {\bf 4}, 251 (1973);
  T.~Kodama,
  Phys.\ Rev.\  D {\bf 18}, 3529 (1978);
  G.~Cl{\'e}ment,
  Gen.\ Rel.\ Grav.\  {\bf 13}, 763 (1981).
\bibitem{hv1997}
D.~Hochberg and M.~Visser,
Phys. Rev. D{\bf 58}, 044021 (1998).
\bibitem{negative}
D.~Ida and S.A.~Hayward, 
Phys.Lett. {\bf A260}, 175 (1999); 
M.~Visser, S.~Kar, and N.~Dadhich,
Phys. Rev. Lett. {\bf 90}, 201102 (2003);
C.J.~Fewster and T.A.~Roman,
Phys. Rev. D{\bf 72}, 044023 (2005); 
P.K.F.~Kuhfittig,
Phys. Rev. D{\bf 73}, 084014 (2006); 
O.B.~Zaslavskii, 
Phys. Rev. D{\bf 76}, 044017 (2007). 
\bibitem{TC}
  J.~L.~Friedman, K.~Schleich, and D.~M.~Witt,
  Phys.\ Rev.\ Lett.\  {\bf 71}, 1486 (1993)
  [Erratum-ibid.\  {\bf 75}, 1872 (1995)];
%
  G.~J.~Galloway, K.~Schleich, D.~M.~Witt, and E.~Woolgar,
  Phys.\ Rev.\  D {\bf 60}, 104039 (1999).
\bibitem{bd}
A.G.~Agnese and M.~La Camera,
Phys. Rev. D{\bf 51}, 2011 (1995);
  L.~A.~Anchordoqui, S.~E.~Perez Bergliaffa and D.~F.~Torres,
  Phys.\ Rev.\  D {\bf 55}, 5226 (1997);
  K.~K.~Nandi, B.~Bhattacharjee, S.~M.~K.~Alam, and J.~Evans,
  Phys.\ Rev.\  D {\bf 57}, 823 (1998);
  P.~E.~Bloomfield,
  Phys.\ Rev.\  D {\bf 59}, 088501 (1999);
  K.~K.~Nandi,
  Phys.\ Rev.\  D {\bf 59}, 088502 (1999);
  K.~K.~Nandi, A.~Islam, and J.~Evans,
  Phys.\ Rev.\  D {\bf 55}, 2497 (1997);
K.K.~Nandi and Y.-Z.~Zhang,
Phys. Rev. D {\bf 70}, 044040 (2004); 
A.~Bhadra and K.~Sarkar,
Mod. Phys. Lett. {\bf A20}, 1831 (2005);
E.F.~Eiroa, M.G.~Richarte, and C.~Simeone,
Phys. Lett. {\bf A373}, 1 (2008).
\bibitem{gb}
D.~Hochberg,
Phys. Lett. {\bf B251}, 349 (1990); 
H.~Fukutaka, K.~Tanaka and K.~Ghoroku,
Phys. Lett. {\bf B222}, 191 (1989); 
D.H.~Coule and K.-i.~Maeda,
Class. Quant. Grav. {\bf 7}, 955 (1990); 
K.~Ghoroku and T.~Soma,
Phys. Rev. D{\bf 46}, 1507 (1992); 
N.~Furey and A.~DeBenedictis,
Class. Quant. Grav. {\bf 22}, 313 (2005); 
F.S.N.~Lobo,
Class. Quant. Grav. {\bf 25}, 175006 (2008).
\bibitem{vkd2003}
M.~Visser, S.~Kar, and N.~Dadhich,
Phys. Rev. Lett. {\bf 90}, 201102 (2003).
\bibitem{thin} 
E.~Poisson and M.~Visser,
Phys. Rev. D {\bf 52}, 7318 (1995);
M.~Ishak and K.~Lake,
Phys. Rev. D {\bf 65}, 044011 (2002);
E.F.~Eiroa and G.E.~Romero,
Gen. Rel. Grav. {\bf 36}, 651 (2004);
F.S.N.~Lobo and P.~Crawford,
Class. Quant. Grav. {\bf 21}, 391 (2004);
F.S.N.~Lobo,
Phys. Rev. D {\bf 71}, 124022 (2005);
E.F.~Eiroa and C.~Simeone,
Phys. Rev. D {\bf 76}, 024021 (2007);
E.F.~Eiroa,
Phys. Rev. D {\bf 78}, 024018 (2008);
J.P.S.~Lemos and F.S.N.~Lobo,
Phys. Rev. D {\bf 78}, 044030 (2008).
\bibitem{sh2002} 
H.-a.~Shinkai and S.A.~Hayward,
Phys. Rev. D {\bf 66}, 044005 (2002).
\bibitem{a-p2002} 
C.~Armendariz-Picon,
Phys. Rev. D {\bf 65}, 104010 (2002).
\bibitem{ggs2008} 
J.A.~Gonzalez, F.S.~Guzman, and O.~Sarbach,
Class. Quant. Grav. {\bf 26}, 015010 (2009);
Class. Quant. Grav. {\bf 26}, 015011 (2009).
\bibitem{bg2001} 
K.A.~Bronnikov and S.~Grinyok,
Grav. Cosmol. {\bf 7}, 297 (2001);
K.A.~Bronnikov and S.V.~Grinyok,
Grav. Cosmol. {\bf 10}, 237 (2004);
Grav. Cosmol. {\bf 11}, 75 (2005).
\bibitem{ms1964} 
C. W.~Misner and D. H.~Sharp, 
Phys. Rev. {\bf 136}, B571 (1964).
\bibitem{roberts1989}
M.D.~Roberts, 
  Gen. Rel. Grav.
  {\bf 21}, 907 (1989). 
\bibitem{sussman1991} 
R.A.~Sussman, 
J. Math. Phys. {\bf 32}, 223 (1991). 
\bibitem{burko1997}
L.M.~Burko, 
Gen. Relat. Grav. {\bf 29}, 259 (1997). 
\bibitem{brady1994} 
P.R.~Brady,
Class. Quant. Grav. {\bf 11}, 1255 (1994). 
\bibitem{ont1994} 
Y.~Oshiro, K.~Nakamura, and A.~Tomimatsu, 
Prog. Theor. Phys. {\bf 91}, 1265 (1994). 
\bibitem{hayward2000} 
S.A.~Hayward, 
Class. Quant. Grav. {\bf 17}, 4021 (2000).
\bibitem{ch2001} 
G.~Clement and S.A.~Hayward,
Class. Quant. Grav. {\bf 18}, 4715 (2001).
\bibitem{gb1967}
I.I.~Gutman and R.M.~Bespal'ko,
Sbornik Sovrem. Probl. Grav. Tbilissi, {\bf 1}, 201 (1967).
\bibitem{madsen1988}
M.~S.~Madsen
Class. Quant. Grav. {\bf 5}, 627 (1988).
\bibitem{hayward1994} 
S.A.~Hayward, 
Phys. Rev. D {\bf 49}, 6467 (1994).
\bibitem{hayward1999}
S.A.~Hayward,
Int. J. Mod. Phys. D{\bf 8}, 373 (1999).
\bibitem{hv1998}
D.~Hochberg and M.~Visser,
Phys. Rev. D{\bf 58}, 044021 (1998).
\bibitem{mhc2009}
H.~Maeda, T.~Harada, and B.J.~Carr,
e-Print: arXiv:0901.1153 [gr-qc]. 
\bibitem{bi1991} 
C.~Barrabes and W.~Israel,
Phys. Rev. D {\bf 43}, 1129 (1991). 
\bibitem{Poisson}
E.~Poisson, 
{\it A Relativist's Toolkit}
(Cambridge University Press, Cambridge, England, 2004).
\bibitem{fj2004}
A.~Feinstein and S.~Jhingan,
Mod. Phys. Lett. {\bf A19}, 457 (2004).
\bibitem{Tipler}
F.J.~Tipler, 
Phys. Rev. Lett. A {\bf 64}, 8 (1977).
\bibitem{Krolak}
A. Kr\'{o}lak, 
J. Math. Phys. {\bf 28}, 138 (1987).
\bibitem{ClarkeKrolak}
C.J.S.~Clarke and K.~Kr\'{o}lak, 
J. Geom. Phys. {\bf 2}, 127 (1985). 
\bibitem{Joshi}
P.S.~Joshi, 
\textit{Global Aspects in Gravitation and Cosmology} (Oxford University Press, New York, 1993).
\bibitem{Clarketext}
C.J.S.~Clarke, 
\textit{The analysis of Space-Time Singularities} (Cambridge University Press, Cambridge, 1993).
\bibitem{kh2004} 
S.A.~Hayward and H.~Koyama,
Phys. Rev. D {\bf 70}, 101502(R) (2004);
H.~Koyama and S.A.~Hayward,
Phys. Rev. D {\bf 70}, 084001 (2004).
\bibitem{geroch1967}
R.P.~Geroch,
J. Math. Phys. {\bf 8}, 782 (1967).
\bibitem{frolov1997} 
A.V.~Frolov,
Phys. Rev. D {\bf 56}, 6433 (1997);
Phys. Rev. D {\bf 59}, 104011 (1999);
Phys. Rev. D {\bf 61}, 084006 (2000).
\bibitem{wesson1978}
P.S.~Wesson,
J. Math. Phys. {\bf 19}, 2283 (1978).
\bibitem{krasinski}
A.~Krasi\'{n}ski, 
{\it Inhomogeneous Cosmological Models}
(Cambridge University Press, Cambridge, England, 1997).
\bibitem{brady1995} 
P.R.~Brady,
Phys. Rev. D {\bf 51}, 4168 (1995). 
\bibitem{application} 
A.~Ishibashi and A.~Hosoya,
Phys. Rev. D {\bf 60}, 104028 (1999);
U.~Miyamoto and T.~Harada, 
Phys. Rev. D {\bf 69}, 104005 (2004);
T.~Harada and H.~Maeda, 
Class. Quant. Grav. {\bf 21}, 371 (2004).





\end{thebibliography}
\end{document}